# A Generalized Framework for Assessing Equity in Ground Transportation Infrastructure: An Exploratory Study

Zahra Halimi, S.M.ASCE,[1] Mohammad SafariTaherkhani, S.M.ASCE,[2] and

Qingbin Cui, A.M.ASCE[3]

[1]Department of Civil and Environmental Engineering, University of Maryland at College Park, MD 20740; Email: zhalimi@umd.edu

[2]Center for Advanced Transportation Technology (CATT), University of Maryland at College Park, MD 20740; Email: msafari@umd.edu

[3]Department of Civil and Environmental Engineering, University of Maryland at College Park, MD 20742; Email: cui@umd.edu

**ABSTRACT**

Ground transportation infrastructure significantly impacts community connectivity, economic growth, and access to essential services such as jobs, education, and healthcare. However, in practice, these infrastructures do not provide equitable services for all, and historical disparities have led to inequitable conditions for many individuals. This paper explores diverse definitions of equity relevant to ground infrastructures, examines the consequences of inequitable transportation systems throughout the history of the U.S. highway system, and explores various approaches for conducting equity analysis. Based on the collected information, a generalized framework is introduced to analyze the impact of these infrastructures on transportation equity. The study also provides a novel equity index that can be used to assess equity considering accessibility and affordability. finally, the framework is applied to a case study in Baltimore City, Maryland, examining the equitable distribution of electric vehicle (EV) chargers.  By demonstrating its practical application, the paper offers managers and policymakers a concise step-by-step approach to analyze transportation equity. This method assesses both the socioeconomic features of affected populations and the distribution of services, contributing to the development of a more sustainable ground transport system.

## INTRODUCTION

Ground infrastructure refers to 'transportation that is conducted over land, rather than on water or in the air.' The network of physical structures such as roadways and a variety of associated elements such as bridges, tunnels, rest areas, gas stations, and systems designed to support efficient and safe movement of vehicles in this infrastructure (CMAC, 2022; Woldesenbet et al., 2016). These infrastructures play a vital role in connecting communities, promoting economic growth, reducing poverty, improving access to education, and supporting sustainable development. However, the distribution and quality of transportation networks have historically shown inequity, leading to disparities in connectivity, mobility, and socio-economic opportunities (Kaiser & Barstow, 2022). The presence of ground infrastructure can contribute to economic prosperity, but it can also place a disproportionate burden on marginalized and underserved communities.

In the U.S., the Infrastructure Investment and Jobs Act (IIJA) allocates substantial funds specifically targeted at improving transportation infrastructure in underserved and marginalized communities (NAPA, 2021; The White House, 2021). This increased attention underscores the need to answer the question: who and what should be considered when addressing equity (Cantilina et al., 2021)? The growing emphasis on social justice in the public sphere, along with existing transportation equity policy requirements, demonstrates that it is the responsibility of transportation practitioners and infrastructure managers to address the inequities faced by vulnerable and historically marginalized populations (Nadimi et al., 2023).

In the literature, equity often refers to the fair distribution of resources (Lewis et al., 2021). However, this term is somewhat amorphous and requires comparison to existing definitions of equity such as universal service or social equity. This definition lays the foundation for developing approaches to infrastructure managers to evaluate the potential and existing equity impacts of transportation planning.

This study offers a detailed examination of equity definitions in the US transportation system and a critical review of key justice concepts such as universal service and social equity by revisiting the historical development of U.S. ground infrastructure and related legislation. Section 2 contextualizes current equity challenges within a broader legislative and historical case study, offering a deeper understanding of how past decisions continue to negatively affect communities. Section 3 discusses various equity measurement methods and their limitations. Section 4 introduces a new generalized framework designed to assess equity in-ground infrastructure. This framework not only considers the socioeconomic status of communities but also includes the dimensions of inequity that such infrastructure imposes on these communities. Section 5 of the paper details the use of a new framework for conducting an equity analysis of electric vehicle (EV) chargers, featuring a novel equity index designed specifically for assessing EV infrastructure in urban settings, exemplified by its application in Baltimore City. This development represents a methodological advancement in quantifying and addressing equity within ground transportation infrastructures. The paper concludes in Section 6 with recommendations for advancing equity in transportation infrastructure, providing key insights for policymakers and managers. This study contributes to the management in engineering domain by providing a comprehensive framework that integrates equity considerations into the decision-making process. Additionally, this framework includes a novel equity index specifically designed for evaluating EV infrastructure, offering a new tool for researchers and policymakers to address equity in urban transportation planning. This framework serves as a vital tool for engineering managers, enabling them to incorporate equity considerations into their project planning and evaluation processes.

**LITERATURE REVIEW**

**Introduction to the Definition of Transportation Equity**

The concept of transportation equity has been a matter of debate for recent decades. This is because equity is a wide-reaching concept rooted in the study of ethics and morality. Many studies define equity as the idea of fairness in distributing services among different populations ([Heydari et al., 2024](); [Lewis et al., 2021]()).

The term equity is used across various fields including politics, economics, and environmental science. As a result, its exact meaning changes depending on the context (Guo et al.,2020). In the United States equity emerged with the universal service concept as a cornerstone of the Communications Act of 1934. Universal service aims to ensure that essential services and infrastructure are accessible to all individuals, regardless of their geographic location, socioeconomic status, or other potential barriers (Choné et al., 2000). In transportation, equity issues first emerged with the Civil Rights Act of 1964, which requires federal agencies to distribute federal resources in the fairest and least discriminatory manner (Welch & Mishra, 2013).

Within the existing body of literature, multiple studies emphasize considering social equity principles in developing the built environment, particularly in light of historical disparities in the USA's ground transport system. However, various perspectives and ideologies exist regarding the precise meaning of social equity within society (Fan & Machemehl, 2011). Social equity is a broad concept that encompasses the fair and just distribution of resources, opportunities, and benefits across all segments of society. It aims to address and rectify disparities that arise due to various factors such as race, gender, and socioeconomic status (Litman, 2022).

Similarly to universal service and social equity, transportation equity aims to ensure all individuals have access to transportation services regardless of socioeconomic status and location. However, transportation equity can be broadened. For instance, transportation equity considers that users of different modes of transportation should receive equitable service. For example, while building a road can provide positive service for vehicle users, if the road lacks suitable facilities for pedestrians and bicyclists, it fails to provide equitable transportation service for all users (Halimi et al., 2024). The distinction between equity and equality lies in their approach to distributing resources or opportunities. While equality involves providing the same resources to all individuals or groups, equity recognizes that individuals have different circumstances and allocates resources based on their specific needs to achieve fair outcomes (Just Health Action, 2010).

Additionally, while transportation inequity is often linked with social disparity, it persists even in predominantly white communities in the U.S., highlighting its broader scope beyond racial lines. The IIJA establishes a rural surface transportation grant program that focuses on increasing connectivity, stimulating economic growth, and improving the quality of life in these communities (NAPA, 2021).

Transportation equity encompasses various elements, including public infrastructure and services, costs and benefits for users, quality of service, external and economic impacts, as well as regulation and enforcement (Litman, 2022). Various definitions have been defined for transportation equity, such as horizontal equity and vertical equity. Horizontal equity ensures equal opportunities and resources among different people and groups with similar needs and abilities (Delbosc & Currie, 2011). This definition means equal costs, benefits, and opportunities for each individual or group. Conversely, vertical equity means allocating resources between different people or groups with diverse societal needs and abilities. It promotes an uneven allocation of resources to reduce disparities among socio-economic groups or individuals, aiming to bridge the gap (Behbahani et al., 2019; Gandy et al., 2023; Litman, 2022). In this paper, we stick to the definitions of horizontal equity and vertical equity for considering equity in transportation.

**Key U.S. Transportation equity related Legislation**

To gain a better understanding of current transportation equity in the US, it is essential to examine the history of equity in related legislation. The development of the road system in the United States represents a monumental evolution from early dirt paths to the sophisticated interstate ground transport networks we traverse today. This transformation was significantly influenced by pivotal legislation such as the Federal Aid Road Act of 1916, which introduced federal funding for road construction, aiming to improve postal road services and general transportation efficiency (Kirk, 2019).

In 1949, Texas established the Farm to Market Road System, a proactive approach to enhance economic equity by improving access from rural areas to urban markets. This system has been crucial in supporting rural economies, reducing the cost of transportation, and increasing the accessibility of agricultural products to major markets, which in turn has bolstered the broader state economy (Prozzi & Harrison, 2007). Later, the Federal-Aid Highway Act of 1956 was a landmark that created the Interstate Highway System, drastically altering American landscapes, economies, and urban configurations. While these developments were pivotal in enhancing national connectivity and economic growth, they also set the stage for several equity challenges, particularly in urban environments where ground transportation such as highways, bridges, and tunnels often divided communities and disrupted local and minority economies (Archer, 2020).

The Civil Rights Act of 1964, particularly Title VI, prohibits discrimination based on race, color, or national origin in programs and activities receiving federal financial assistance, including transportation. Despite this legislation, there remains a need for a more equitable transportation system in the USA (Guo et al., 2020; Welch & Mishra, 2013). In 2021, the U.S. Congress passed the IIJA, a comprehensive bill addressing nationwide infrastructure needs with substantial investments across various sectors. This act aims to rebuild America's roads, bridges, and railways, focusing on communities that have often been left behind (NAPA, 2021).

**Historical Discrimination, Equity Challenges, and Impacts on U.S. Ground Infrastructure**

In the past, the assessment of transportation system performance focused heavily on trip speeds, giving preference to faster but more costly modes like driving while neglecting the more affordable options such as walking, cycling, and public transportation. Consequently, the outcomes of these decisions often proved to be unfair. For instance, various ground transport projects resulted in the deterioration of mixed-use urban areas, as the planning process prioritized the benefits for motorists without adequately addressing the adverse effects on accessibility, liveability, and economic opportunities for residents. While some of these infrastructures are now facing criticism and potential removal, the damage caused remains irreversible. However, the implications of these equity issues have frequently been disregarded or overlooked (Feng & Wu, 2003; Litman, 2022). It is crucial to recognize that projects addressing historical discrimination differ significantly from general transportation improvements. These projects specifically aim to correct past injustices by ensuring equitable access to transportation services. Table 1 lists examples of ground transport projects in the USA and highlights their significant impacts on community equity. It is important to note that while the table identifies predominant equity issues for each project, other related equity issues may also be present due to their interconnected nature.

**Table 1 Overview of Equity Issues in U.S. Ground Transport Projects**

| Project Name | Location | Year of Completion | Equity issue |
|---|---|---|---|
| Highway 99 Viaduct | Seattle, WA | 1953 | Community Isolation, Restricted Accessibility |
| Staten Island Expressway (I-278) | Staten Island, NY | late 1950s and early 1960s. | Environmental and Health Impact |

| I-35 through East Austin | Austin, TX | 1959 | Community Division, Economic and Social Displacement |
| I-95 through Overtown | Miami, FL | 1960s | Community Division, Economic and Social Displacement |
| Sheridan Expressway | Bronx, NY | 1962 | Community Isolation, Restricted Accessibility |
| I-375 | Detroit, MI | 1962 | Community Division, Economic and Social Displacement |
| I-81 | Syracuse, NY | 1969 | Environmental and Health Impact |
| I-70 through Denver | Denver, CO | 1960s and 1970s, | Community Isolation, Restricted Accessibility |
| Kensington Expressway | Buffalo, NY | 1971 | Economic and Social Displacement, Restricted Accessibility |
| Franklin-Mulberry Expressway | Baltimore, MD | 1979 | Community Division, Economic and Social Displacement |

Ground transportation infrastructures such as highways and expressways often create significant inequities by impacting urban landscapes and community accessibility. In New York, the Sheridan Expressway and Denver's I-70 exemplify how expressways can serve as barriers that restrict access to essential services and disrupt community connectivity (Goetz, 2023; Mohl, 2012). Similarly, the Highway 99 Viaduct in Seattle divided the city, severing the waterfront from downtown (WSDOT, 2023). In Buffalo, the construction of the Kensington Expressway destroyed Humboldt Parkway, leading to community isolation, population decline, and racial segregation (Andrew Emanuele & Katz).

Projects like I-35 through East Austin and I-95 through Overtown highlight significant issues of community division and socio-economic displacement. These highways, constructed through primarily minority neighborhoods, caused major displacement and prolonged economic decline (Archer, 2020; Murdoch, 2020). Similarly, the Franklin-Mulberry Expressway's development in the 1970s devastated West Baltimore's Black communities, demolishing 20 city blocks and displacing 971 households, along with businesses, churches, and schools (TRANSPORTATION, 2022). Such divisions not only perpetuate socio-economic disparities but also erode the social fabric of communities.

On the environmental front, highways like I-81 in Syracuse and I-278 in New York brought additional challenges, including increased pollution and associated health risks to nearby residents, highlighting the environmental and public health consequences of urban transport projects (Meyer & Dallmann, 2022; Teron, 2022).

It should also be noted that the consequences of inequity in ground infrastructure are often interconnected, and projects developed during the expansion of the interstate highway system in the USA share similar characteristics. These include the displacement of people and businesses, which often leads to economic burdens, community isolation, and environmental issues.

## REVIEWING APPROACHES FOR EQUITY ANALYSIS: METHODS AND RELATED RESEARCH

This paper examines a range of studies that propose different methods for assessing equity in transportation infrastructure. These methods can be classified into five groups, which encompass Cost-Benefit analysis (CBA), mismatch analysis (e.g., utilizing GIS mapping), equity indices or measurements regression modeling like multivariate regression, and optimization modeling (Table 2). Each equity assessment method will be thoroughly discussed in the subsequent subsections.

**Cost-Benefit Analysis.** Cost-benefit analysis (CBA) has been widely employed to assess infrastructure investments, yet it possesses certain limitations. Notably, it falls short in addressing equity considerations and tends to undervalue the long-term environmental implications through its discounting mechanism (Jones et al., 2014; Ludwig et al., 2005). When contemplating the guidelines for Cost-Benefit Analysis employed by governmental agencies, the paramount benefit for a project is often the value of time saved. The monetary valuation of travel time has been a subject of critical discussion in Benefit-Cost Analysis (Hensher, 2001). For instance, Specifically, investing in bus services for marginalized communities might not yield as much financial return as other transportation modes that serve a broader demographic (Levine, 2013). Furthermore, the time-saving valuation fails to account for differences in trip frequency across socio-economic statuses. Higher socio-economic groups typically benefit more from transportation improvements due to more frequent travel, whereas lower socio-economic groups may continue to experience accessibility issues (Martens & Di Ciommo, 2017).

Including equity in cost-benefit analysis (CBA) presents challenges due to the need to quantify complex, intangible social impacts and the typical focus on aggregate benefits, which may obscure uneven distribution among societal groups. Scientists can help by developing estimation methods that incorporate distributional weights, although this is complicated by the need for societal value judgments (Hahn, 2021). Another critical concern is choosing an appropriate discount rate for CBA. The discount rate significantly impacts project efficiency assessment and intergenerational equity effects. Proper consideration of the discount rate ensures a balanced evaluation of long-term benefits and impacts across generations (Di Ciommo et al., 2014; Di Ciommo & Shiftan, 2017).

**Mismatch analysis.** Mismatch analysis using basic descriptive statistics is a widely used approach to assess the equity performance of transportation systems, originating from early research evaluating public transport effectiveness among various demographic groups (Currie, 2004). This technique usually entails displaying cost/benefit data in maps or tables for visual comparison of distributions, offering a clear and intuitive understanding of equity performance across different zones or communities (Guo et al., 2020). Statistical metrics for a zone typically include average values, such as exposure to transportation burdens like traffic congestion, air, and noise pollution, and may also encompass the median, maximum, minimum, and standard deviation (Bills & Walker, 2017; Rodier et al., 2009).

This straightforward mapping method effectively presents detailed macroscopic information making it capable of evaluating both horizontal and vertical equity. For horizontal equity, the selected equity criteria within each project study area must be considered (Golub & Martens, 2014). Likewise, vertical equity can be evaluated by mapping socio-economic characteristics and selected equity measures in two maps, enabling a better understanding of the equity distribution (Kaplan et al., 2014). While mismatch analysis provides intuitive insights based on socio-economic characteristics and selected equity criteria, It lacks quantitative measures or performance indicators and also falls short in comparing the distribution of equity status (Guo et al., 2020).

**Equity Index Analysis.** Using the equity index overcomes the limitations of previous methods that lacked quantitative measures, providing a measure for comparing the equity levels across different communities (Guo et al., 2020; Jafari et al., 2020). Defining equity index is common in fields such as economy, social science, and public health. In the study on equity in the workforce, the equity index is defined using factors such as ethnic and racial diversity, gender diversity, corporate policies, equitable pay, and transparent recruitment. This index provides a structured method to measure and improve equity within organizations (Karakhan et al., 2021).

In transportation studies, the Atkinson index is used to measure inequality in the distribution of resources or benefits, with values ranging from 0 (perfect equality) to 1 (complete inequality). Its key feature is an "inequality aversion" parameter that reflects societal values and priorities, allowing for customization based on different concerns about inequality. Moreover, the index can be broken down into subgroups, with the aversion parameter adjustable to emphasize the impacts on lower-end individuals (Guo et al., 2020; Levy et al., 2006).

The Comparative Environmental Risk Index (CERI) is a standardized measure for assessing environmental justice and evaluating the risk exposure of different demographic groups, including non-whites and those with low socio-economic status. A CERI value above 1 indicates higher exposure of non-white people to environmental risks compared to white people, and vice versa. This index can also be applied to other subgroups, such as individuals with low socio-economic status (Harner et al., 2002).

The Social Vulnerability Index (SVI), which incorporates factors like age, poverty rate, and unemployment rate, is commonly used to address equity. By combining equity-related features and the SVI index, the equity index can assess both horizontal equity (comparing similar groups) and vertical equity (Li et al., 2023). While the equity index approach offers quantitative assessments of equity and accounts for subgroups in vertical equity analyses, it has notable disadvantages. Firstly, an index may not cover all dimensions of equity, often focusing only on specific aspects, and its complexity can make it more challenging to implement compared to simpler methods. Additionally, the ability to customize parameters such as the inequality aversion factor can introduce biases if not aligned with broader societal values, potentially skewing results and leading to inaccurate representations of equity levels.

**Regression Modelling.** Another approach to assess the equity measure for a community is to use regression analysis to examine the relationship between different socio-economic characteristics of people and equity criteria. Various methods can be used to assess vertical transportation equity, including ANOVA, correlation analysis, and regression modeling. One-way ANOVA and bivariate correlation analysis are simple ways to examine the link between cost/benefit measures and social and demographic factors like age. Regression analysis, considering multiple factors like household income and education level, offers a more comprehensive approach, indicating the relationship direction and strength through the coefficient sign and magnitude (Guo et al., 2020).

Regression methods are effective for studying vertical equity as they can depict the relationship between transportation costs/benefits and socio-demographic factors. Additionally, basic statistical measures of variation in the univariate distribution of cost/benefit variables between geographic areas can provide insights into horizontal equity (Guo et al., 2020). For instance, a study assessed bridge conditions across various community demographics using the *pglm* package for R and linear regression. This analysis explored correlations between bridge conditions and socio-economic factors, effectively measuring equity by evaluating disparities in infrastructure quality across diverse communities (Gandy et al., 2023).

Utilizing regression analysis in transportation equity studies offers a robust method to dissect and understand the dynamics at play. However, careful consideration must be given to the design, implementation, and interpretation of these models to avoid common pitfalls and ensure the results are both reliable and actionable.

**Optimization Modelling.** Optimization models are essential tools in transportation planning, offering significant potential to assess and enhance transportation equity. These models aim to optimize specific objectives such as minimizing travel time, reducing costs, or maximizing service coverage while adhering to various constraints like budget limits or geographical barriers. The choice of optimization technique depends on model complexity. Validation, sensitivity analysis, and trade-off evaluation are crucial. The results should be interpreted and implemented to inform decision-making. The specific details of the optimization model will depend on the context and objectives of the equity analysis (Behbahani et al., 2019).

Halimi, SafariTaherkhani, and CuiFor instance, a bi-level optimization model proposed to address the PTNRP-AEI problem incorporates spatial equity considerations. This model aims to minimize overall costs in the higher-level sub-problem, considering user, operator, and unmet demand costs while adhering to various constraints. The lower-level sub-problem optimizes user routing in the network to minimize travel costs and transfers for transit users. The bi-level PTNRP-AEI optimization model employs a Genetic Algorithm (GA) meta-heuristic approach for a solution. However, a limitation is its failure to address social (vertical) equity concerns. Social equity involves the fair treatment of diverse individuals and groups, ensuring the equitable distribution of costs and benefits based on their unique needs and abilities (Feng & Wu, 2003) .

Optimization algorithms offer the advantage of effectively balancing diverse transportation needs by optimizing various constraints and functions, making it crucial to carefully define these parameters to accurately measure and enhance transportation equity.

**Table 2 Approaches for Analyzing Social Equity in Transportation Infrastructure**

| Approach | Advantages | limitation | Researcher(s) |
|---|---|---|---|
| Cost-Benefit Analysis | - Widely accepted method<br>- Enables Direct Comparisons | - Focus on aggregated benefit<br>- Difficulty in quantifying the equity impact<br>- Sensitivity to discount rate | (Di Ciommo et al., 2014; Di Ciommo & Shiftan, 2017; Hahn, 2021; Martens & Di Ciommo, 2017) |
| Mismatch analysis | - Simple method (Intuitive Visualization)<br>- Consider both horizontal and vertical equity | - Lack of providing quantitative measures<br>- Inadequate Comparison Capabilities | (Bills & Walker, 2017; Kaplan et al., 2014; Rodier et al., 2009) |

| Approach | Advantages | limitation | Researcher(s) |
|---|---|---|---|
| Equity Index Analysis | - Quantitative indicator to measure the equity<br>- Consider both horizontal and vertical equity | - Limited Scope<br>- Complex Implementation<br>- Potential for Bias | (Behbahani et al., 2019; Harner et al., 2002; Karakhan et al., 2021; Levy et al., 2006; Palmateer & Levinson, 2017) |
| Regression | - Examine the link between equity measures and socio-demographic characteristic<br>- Consider both horizontal and vertical equity | - Sensitivity to Outliers<br>- Interpretation Complexity | (Cuthill et al., 2019; Gandy et al., 2023; Guo et al., 2020) |
| Optimization Modeling | - Consideration of decision variables, objective functions, constraints, and equity considerations<br>- Suitable for large scale area | - Sensitive to algorithms, variables, and constraints.<br>- Complexity | (Feng & Wu, 2003) (Behbahani et al., 2019) |

**PROPOSED GENERAL FRAMEWORK FOR EQUITABLE INFRASTRUCTURE**

Based on the literature review, assessing equity in transportation infrastructure involves six main steps. The proposed general framework offers a guideline for different steps of equity analysis considering vertical and horizontal equity in ground transport infrastructure.

**Step 1: Identify the project target group.** The initial analysis requires identifying the individuals affected by the infrastructure. To achieve this, defining a specific unit based on the project and equity analysis type is essential. Spatial unit areas, such as census tracts, census blocks, or traffic analysis zones (TAZs), can be chosen for the analysis to represent the entire project's target population (Guo et al., 2020). In other words, in this step, we should answer the question of who is affected by that transportation infrastructure.

**Step 2: Collecting Socio-economic Data.** This step involves collecting socio-economic and demographic data of individuals or groups affected by transportation infrastructure. The data may include age, gender, marital status, income, expenses, physical abilities, education, job, and other relevant aspects. In transportation planning, grouping might also involve travel-related factors like mode of transportation, vehicle type, purpose of travel, and road conditions (Behbahani et al., 2019; Guo et al., 2020). It is important to understand that excluding the demographics, and resources of neighboring communities from the equity analysis can lead to a misrepresentation of the impact of transportation infrastructure on local communities in the decision-making process (Ahmed & Garvin, 2022; Gandy et al., 2023). By identifying groups with similar features, we can conduct horizontal equity analysis, ensuring equal treatment within these homogeneous groups. Conversely, by considering units with different features, we can conduct vertical equity analysis, addressing disparities between diverse socio-economic and demographic groups.

**Step 3: Identify the project-disadvantaged group.** During this step, the disadvantaged community is identified based on project requirements and the collected data in the previous step. Multiple methods exist to calculate the disadvantaged community using various criteria. For instance, (Manaugh & El-Geneidy, 2012) proposed using a composite index comprising five indicators: median household income, the proportion of locals born abroad, the percentage of individuals with a high school diploma as their highest level of education, the percentage of pedestrians using public transport for work trips, and the availability of low-skill employment requiring only a high school diploma, considering competition from individuals with similar educational backgrounds. All scores from these indicators were standardized (z-score), and socially disadvantaged neighborhoods were identified as the lowest decile neighborhoods. These neighborhoods are characterized by low-income, transit-dependent immigrants with low academic levels, facing linguistic and material limitations, and limited access to suitable job opportunities. It is worth noting that various decision-makers may have different perspectives on disadvantaged groups.

(Lucas, 2012; McCulloch, 2006) highlights that within the European context, communities facing transportation disadvantages are identified through their potential for "social exclusion" due to inadequate transportation access. Rather than labeling entire communities as disadvantaged from the outset, the focus is on individuals and locations at risk of diminished engagement in essential life activities. Conversely, in the United States, historical patterns of racially biased policymaking have led to a concentration of transportation disadvantages among people of color and those with lower incomes. This paper does not aim to argue or advocate for a particular group to be labeled "disadvantaged."

Moreover, there are some tools that can help understand the transportation of disadvantaged communities. For instance, in the U.S., the U.S. DOT Equitable Transportation Community (ETC) Explorer and Areas of Persistent Poverty & Historically Disadvantaged Communities tool, Climate & Economic Justice Screening Tool (CEJST) are web applications designed to provide insights into and explore communities that are underinvested in transportation.

**Step 4: Establish equity criteria in alignment with project goals.** In this step, it is essential to define specific criteria for equity that are directly aligned with the project's objectives. This process involves determining the key factors and considerations that will be used to assess equity in the context of the project's scope and purpose. In the following, there are some primary criteria for assessing equity based on the previous section that reviews the main equity criteria.

*Accessibility.* The accessibility dimension of equitable Transportation infrastructure focuses on ensuring that all individuals, regardless of their socio-economic status, have convenient access to essential services, such as education, healthcare, employment, and social amenities. According to study by (Hansen, 1959), one of the primary references for defining accessibility, is characterized as the potential for interaction, which means that the ability to do daily life with a variety of transportation modes, a range of destination choices, and proximity to these destinations. Within the scope of transportation equity, accessibility has been employed to provide equity for a special group of people. For example, In the U.S, there are initiatives like the U.S. Department of Transportation's Disability

Policy Priorities designed to enhance access for people with disabilities across various areas, thereby promoting transportation equity by facilitating safe and equitable access (US DOT, 2023).

*Connectivity.* Connectivity refers to the abundance of connections within path or road networks and link efficiency. Extensively connected networks have many short links, intersections, and limited dead-ends. Higher connectivity results in shorter travel distances and more route options, facilitating direct travel between destinations. In the context of transportation equity, facilities can sometimes present obstacles to connectivity. This means that the transportation system's infrastructure may hinder direct access to education, employment, and routine destinations. Therefore, establishing equity criteria that consider community connectivity, including mobility, access, or economic development, can be an approach for doing equity analysis (Delbosc & Currie, 2011).

*Affordability.* Affordability means expenses within one's income, covering primary needs within budget. For housing and transportation, costs should not exceed 45% of total income to be considered affordable. For an average family allocating 30% on housing, transportation spending should be limited to 15%. Unfortunately, most households spend more, leading to transportation-related financial burdens, disproportionately affecting low-income households (Litman, 2022).

*External cost.* External costs refer to the costs imposed on society as a whole by transportation activities but are not directly paid for by the users of the transportation system. These costs can include congestion, crash risk, and diseases like asthma, among others (Litman, 2022). Additionally, factors such as poor lighting, restricted visibility, and a lack of clear escape routes contribute to a heightened fear of crime, further exacerbating safety issues by creating an environment that feels insecure (Anciaes & Jones, 2018; Khojastehpour et al., 2022).

*Social Justice and Inclusivity.* Equitable Transportation infrastructure seeks to address social inequalities and promote inclusive mobility. This factor examines Transportation infrastructure's impact on marginalized populations, such as low-income areas, those who are disabled, and people of racial and ethnic minorities.

*Environmental Sustainability.* Equity in Transportation infrastructure extends beyond social and economic aspects and includes environmental considerations. This factor examines sustainable transportation practices, such as promoting public transit, non-motorized transportation, and reducing carbon emissions to ensure equitable and environmentally friendly Transportation infrastructure. Other concerns include rising heat zones, which pose serious risks to both human health and the environment, excessive noise, and offensive odors (Hamersma et al., 2014; van Eldijk et al., 2022).

*Fair share of resources.* Transportation infrastructure should provide equitable benefits and costs to individuals with similar abilities and needs, regardless of their mode of transportation. This means that both drivers and non-drivers should experience comparable levels of safety and accessibility (Litman, 2022). It is noted that Assessing transportation equity is complex due to the various viewpoints and consequences involved. To navigate this complexity effectively, it is often advisable to establish a set of measurable criteria aligned with the specific service types of road infrastructure, social values of people who make decisions and the society they represent (Behbahani et al., 2019; Litman, 2022).

**Step 5: Determine the equity analysis approach.** In this step, the most suitable approach should be selected based on the level of equity analysis and available data. While it is important to acknowledge that there is no universally standardized approach for measuring equity impact of infrastructure assets, several approaches have been proposed over the past few decades for assessing equity in transportation infrastructure (Guo et al., 2020). Based on the literature review, these approaches can be categorized as follows: cost-benefit analysis, mismatch analysis, equity index analysis, regression modeling, or optimization modeling. Moreover, equity criteria are evaluated using various benchmarks, including per trip, per capita, per passenger mile, or dollar. Each benchmark represents different hypotheses and viewpoints. For example, per capita analysis assumes that everyone receives an equal share of resources, while per-mile or per-trip analysis assumes that frequent travelers should receive a larger allocation of public resources (Litman, 2022). Therefore, it is important to select equity criteria benchmarks carefully to ensure the result of equity analysis is aligned with real needs.

**Step 6: Documentation and monitoring of the equity criteria.** The final step is to document and monitor the equity criteria over time. This time-based monitoring allows for a deeper understanding of how equity considerations evolve and fluctuate within the context of the transportation infrastructure project. By regularly documenting and analyzing the equity criteria at various intervals, researchers and policymakers gain valuable insights into the dynamics of equity impacts on different communities and demographic groups.

**CASE STUDY**

This study employed a meso-scale analysis using an Equity Index Analysis to evaluate the equity in the distribution of electric vehicle charging stations across Baltimore, Maryland. Many policy initiatives, such as the Infrastructure Investment and Jobs Act, include investments in EV infrastructure to support widespread adoption. The act allocates $7.5 billion specifically for EV charging stations ([The White House, 2021](#)). Conducting equity analysis for EV chargers as part of ground transport infrastructure is crucial to ensure that all communities, including marginalized and low-income groups, have fair access to this essential resource. Equitable distribution of EV chargers helps reduce emissions, improve air quality, and create economic opportunities, thereby promoting the widespread adoption of electric vehicles (EVs). Furthermore, it aligns with social and regulatory goals, such as those outlined in the Infrastructure Investment and Jobs Act, which emphasize the importance of fairness and justice in infrastructure development ([Esmaili et al., 2024](#)).

**Step 1:** According to the first step of the framework, the unit of analysis should be considered. In this analysis, we selected the census tract as the unit of analysis due to its ability to provide detailed demographic information and its relevance in reflecting neighborhood-level differences in socio-economic factors, transportation needs, and infrastructure. Although much research has focused on optimizing EV charging station (EVCS) locations for maximum EV flow and cost reduction, these studies have not specifically measured the equity of EVCS distribution. Consequently, there is no existing standard or benchmark for evaluating the equity of EVCS distribution that incorporates both accessibility and affordability, which are crucial factors for ensuring fair access to charging infrastructure. Given the absence of studies specifically addressing the equity index of EV charging stations, this study proposes a new measure in terms of equity index that incorporates parameters of accessibility (measured by miles of range per hour of charging) and affordability (cost of charging).

**Step 2:** The analysis required comprehensive information about the available charging stations in each census tract, which included identifying whether the charging station is Level 1, Level 2, or DC fast charging, and the total number of each type of charging station available. This data was sourced from the Alternative Fuels Data Center, a part of the U.S. Department of Energy (DOE) ([Center, 2023](#)). A summary of the charging station distribution for each charging level can be seen in *Figure 1*.*Figure 1*

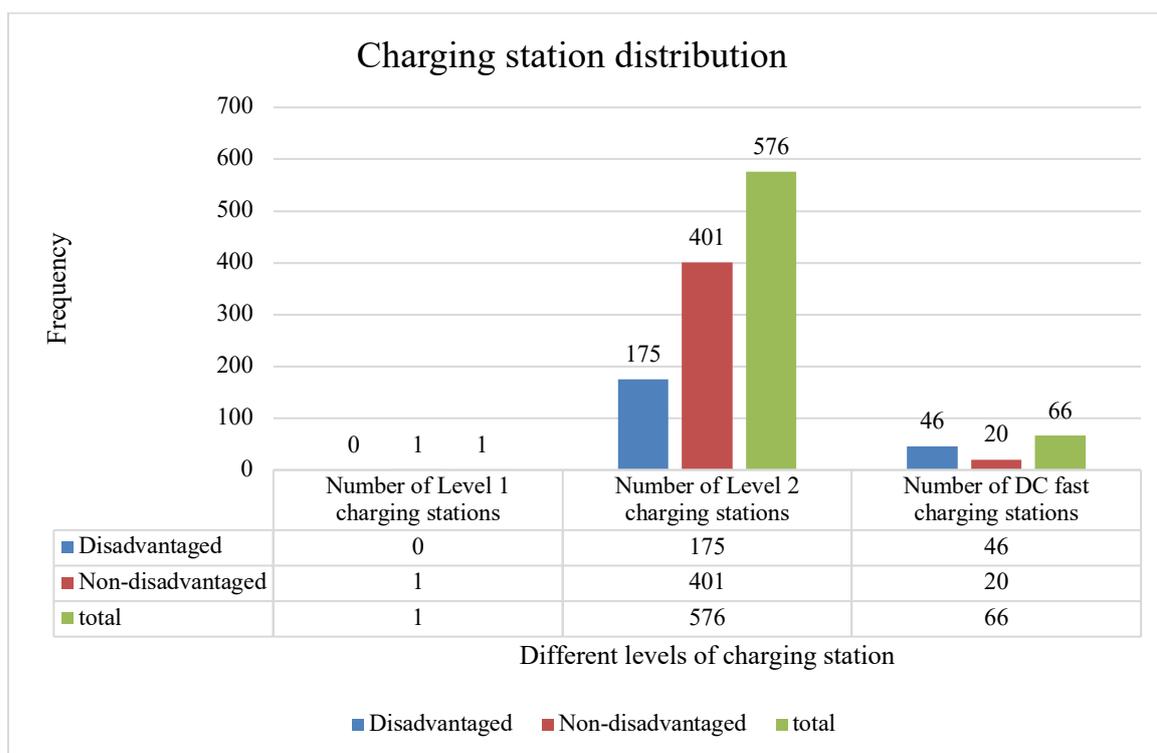

**Figure 1. Summary distribution of different levels of charging stations**

**Step 3:** In addition to previous data, demographic information for each census tract was gathered from the United States Census Bureau (Bureau). These two datasets were then aligned to understand the current situation of charging stations in each census tract. Subsequently, disadvantaged census tracts were determined using the Equitable Transportation Community (ETC) explorer tool provided by the US Department of Transportation (DOT) and linked to the created dataset (*USDOT Equitable Transportation Community (ETC) Explorer*). The utilization of the ETC Explorer tool is primarily driven by its proficiency in providing a comprehensive understanding of transportation disadvantages at the community level. This tool integrates a diverse range of data sources, such as the CDC's Environmental Justice Index (EJI), NOAA's Climate Mapping for Resilience and Adaptation (CMRA), FEMA's National Risk Index, and the EPA's Smart Location Database. These varied sources collectively offer a thorough perspective on the multifaceted impact of transportation across five key dimensions: Transportation Insecurity, Climate and Disaster Risk Burden, Environmental Burden, Health Vulnerability, and Social Vulnerability. By leveraging these components, the tool helps users identify census tracts that are disadvantaged in terms of transportation through a multi-component approach (ETC Explorer Technical Documentation). Each census tract was assigned a composite score based on the above dimensions, and other socio-economic factors such as race, and income. Tracts scoring above a certain threshold were categorized as disadvantaged communities, indicating higher levels of transportation and environmental disadvantages. Tracts scoring below this threshold were categorized as non-disadvantaged communities (*USDOT Equitable Transportation Community (ETC) Explorer*).

In the equity assessment of charging stations in Baltimore, we choose to consider both accessibility and affordability related factors for the equity analysis. Accessibility, which measures the ease of reaching destinations and the availability of activity opportunities, plays a significant role in influencing people's socio-economic activities. By analyzing variations in accessibility levels, we can identify potential inequities arising from unequal access to charging stations (Behbahani et al., 2019; Li et al., 2022). Furthermore, affordability is a vital factor in encouraging the adoption of electric vehicles (EVs), particularly among low-income households and communities (Menghwani et al., 2020). A lack of affordable EV chargers can lead to a situation where only wealthier households can access the environmental and health benefits of EVs, exacerbating existing disparities (Sharma et al., 2023).

**Step 4:** Next, the equity index is chosen for this case study and calculated using a combination of accessibility and affordability for each census tract. The equity index is chosen because it provides a quantifiable measure of the equitable distribution of EV charging stations. The equity index allows us to assess disparities across different census tracts in a meso-scale. This scale is beneficial for understanding the localized impacts of infrastructure distribution. Accessibility is a primary factor as it reflects how easily residents can reach EV charging stations, regardless of socio-economic status. Affordability is also crucial, as high charging costs can be a barrier for low-income households, which limits EV adoption. The miles of range per hour of charging in equations 1 represents the quantified benefit of accessibility, while the charging price represents the quantified cost of affordability. Also, per capita measurements are generally recommended for assessing impacts because equity is concerned with people (Litman, 2022). The ratio is divided by the number of individuals in each census tract to account for population differences. To standardize the equity indicator from 0 to 1, the equity ratio in each census tract is divided by the maximum equity ratio among all the census tracts. The Equity index for each census tract is calculated using the following formula.

Equation 1:

$$E_j = \frac{\sum_i \alpha_i N_{ij} / C_{ij}}{P_j} \bigg/ Max\left(\frac{\sum_i \alpha_i N_{ij} / C_{ij}}{P_j}\right) \forall j$$

Notation:

$E_j =$ Equity index at census tract $j$

$\alpha_i =$ Miles of range per 1 hour of charging at charging station $i$ $\left(\frac{Miles}{hr}\right)$

$i \in (level\ 1, level\ 2, DC\ fast\ charging)$

$N_{ij} =$ Number of charging station type $i$ at census tract $j$

$C_{ij} =$ Cost of charging at charging station type $i$ in census tract $j$ $\left(\frac{\$}{KWh}\right)$

$P_j =$ Population of individuals over 18 in census tract $j$

The calculated index reflects charging station equity measure at each census tract, adjusted for population, accessibility and charging cost, with higher values indicating better equity status. The index considers factors like charging station type, number, cost, and census tract population of individuals over 18. It ranges from 0 to 1, where 0 indicates no equity and 1 represents maximum equity. In the analysis, the equity ratio is calculated by determining the values of α and C, which are based on the types of charging stations and their charging speed and cost ([Alternative Fuels Data Center](#)).

1. Level 1 charging stations ($\alpha_1$) offer a rate of 5 miles of range per 1 hour of charging. These stations typically have a low cost of charging at $0.10 per kilowatt-hour (KWh).

2. Level 2 charging stations ($\alpha_2$) are faster and can provide 25 miles of range per 1 hour of charging. These stations, however, come at a higher cost, charging at $0.18 per KWh.

3. DC fast charging stations ($\alpha_3$) provide the highest charging speed among the three types, offering as much as 300 miles of range per 1 hour of charging. As expected, the cost of charging at these stations is the highest, at $0.36 per KWh. Each of these parameters and their respective units play crucial roles in the computation of the accessibility index in each census tract.

**Steps 5 and 6:** The analysis results revealed an unexpected heterogeneity within both disadvantaged and non-disadvantaged communities. As shown from the results in *Figure 2*, the majority of non-disadvantaged and disadvantaged tracts have an accessibility index of 0.1, representing relatively low access to charging stations. However, some differences are observed between the two categories:

- The average accessibility index for non-disadvantaged community census tracts is approximately 0.19, which is higher than that of disadvantaged community census tracts, which have an average accessibility index of approximately 0.14.
- However, the disparity is not substantial, indicating that charging station accessibility is relatively similar in both types of tracts.
- For non-disadvantaged community tracts, there are some tracts with high accessibility indices. A slight portion (about 13%) of these tracts have an accessibility index of 0.3 or more. One tract also reached the highest accessibility index of 1.
- For disadvantaged community tracts, the distribution of higher accessibility indices is slightly less than in non-disadvantaged tracts, but three tracts still have an accessibility index of 0.5.
- Notably, the highest index within these tracts is 0.9.

The results suggest that although non-disadvantaged tracts, on average, have slightly better access to charging stations, the overall distribution of these stations is not ideal in either category. The majority of tracts in both categories have relatively low access, and the number of tracts with high accessibility is minimal. These findings indicate a need for more equitable distribution and better accessibility to charging stations in both non-disadvantaged and disadvantaged census tracts. While there is a slight disparity favoring non-disadvantaged tracts, the results suggest that all tracts, regardless of their socio-economic status, could benefit from improved charging station accessibility.

The low equity index emphasize the challenge of insufficient EV charging infrastructure in both disadvantaged and non-disadvantaged communities. This observation aligns with findings from other studies ([Brown et al., 2024](#); [Loni & Asadi, 2023](#)). As of January 2023, the US reported approximately

*Halimi, SafariTaherkhani, and Cui*

51,228 EV charging stations, marking a ratio of 18.5 EVs per charging port, significantly higher than the international standard of 12 to 1 (Loni & Asadi, 2023; Outlook, 2022; USDOE, 2024). In California, this ratio is about 27 to 1, highlighting a critical gap that needs addressing, especially with the expected increase in EVs due to new legislation (Brown et al., 2023). This widespread issue supports our results and emphasizes the need for a strategic enhancement of the EV charging network to ensure equitable access.

*Figure 3* illustrates the spatial distribution of the Equity Index in Non-Disadvantaged and Disadvantaged Census Tracts that have at least one EV charger station. Disadvantaged communities are represented in shades of yellow, and Non-disadvantaged communities are represented in shades of blue. The map reveals a spatial heterogeneity within both sets of communities. Central Business Districts (CBD) within non-disadvantaged communities exhibit a concentration of higher accessibility values, reflecting an imbalance where more affluent areas have greater access to charging infrastructure. Spatial heterogeneity is evident within both community types, with a varied distribution of Equity Index scores across tracts. Additionally, it is evident from *Figure 2* and *Figure 3* that one census tract has an unexpectedly high equity index. The I-895 highway passes through this census tract, which can contribute to the high equity index of this disadvantaged community.

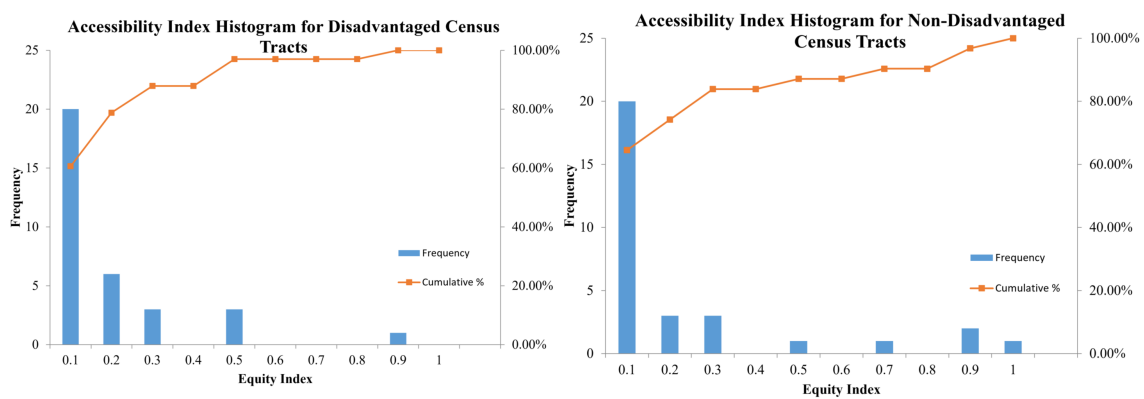

**Figure 2. Equity index histogram for disadvantaged and non-disadvantaged census tracts**

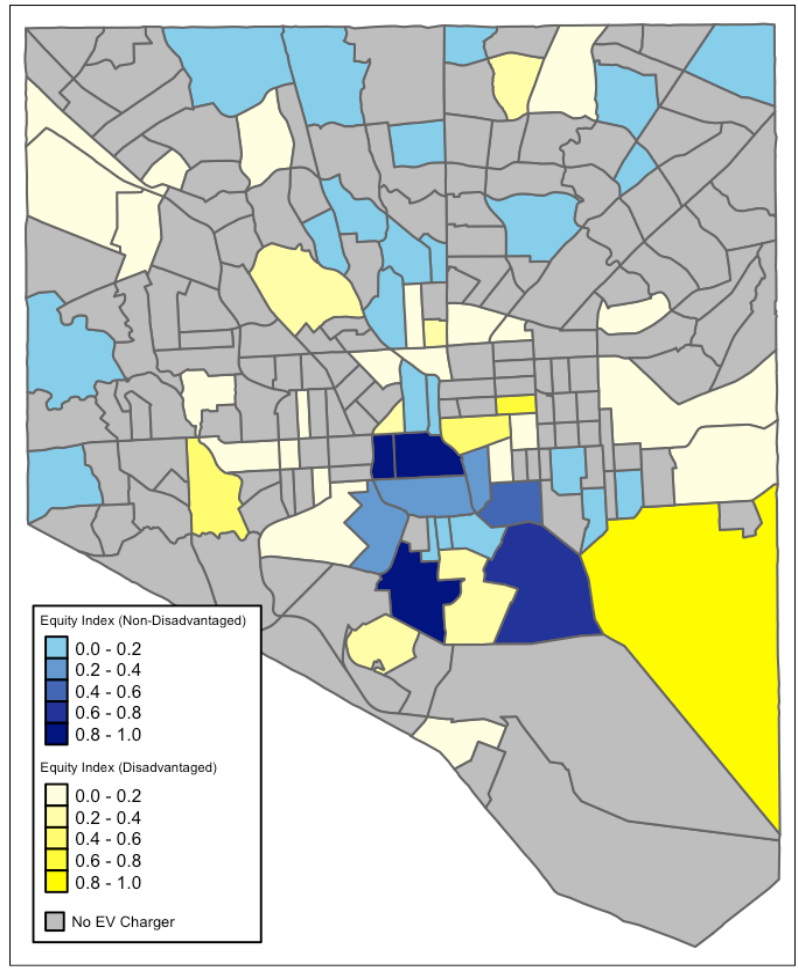

**Figure 3. Distribution of equity index in Non-Disadvantaged and Disadvantaged census tracts**

These results hold valuable implications for both private sector stakeholders and the U.S. Department of Transportation. These results provide a nuanced picture of the current state of charging station distribution, highlighting areas that are in immediate need of infrastructure expansion. This data-driven approach can guide efforts to expand the charging network in a manner that addresses these discrepancies, thereby promoting equity in the accessibility to charging infrastructure. It helps in moving towards targeted interventions, ensuring that every community, regardless of their socio-economic status, has adequate access to charging stations.

In light of the rapid growth of charging stations, it is recommended to perform this analysis at short, regular intervals. This will enable an ongoing assessment of the progress of the deployment of charging stations in these census tracts, which will be instrumental in ensuring an equitable distribution of this essential infrastructure.

**Sensitivity Analysis**

The sensitivity analysis presented explores how changes in electric vehicle charging infrastructure affect equity across census tracts. Additionally, the efficacy of distribution strategies to improve access to charging stations equitably is investigated.

Sensitivity analysis is a critical component of the study. (1) It allows us to evaluate how changes in key variables, such as charging speed and cost, impact the equity of EV charging infrastructure distribution. (2) sensitivity analysis provides a check for the equity index calculations. By examining how sensitive the results are to variations in the input parameters, we can assess the stability of findings. This increases the validity of recommendations by ensuring that conclusions are independent of specific data points or assumptions. (3) It allows us to test different strategies for increasing equity. For instance, based on findings of the strategies for increasing the equity section, we recommend performing population-proportionate distribution of charging stations, as this approach significantly enhances accessibility.

The sensitivity analysis in ***Figure 4*** illustrates the impact of charging speed (alpha values) and cost on the average Equity Index for both disadvantaged and non-disadvantaged census tracts. The x-axis represents the miles of range per hour of charging ($\alpha$ values) for plots on the left side and cost of charging in terms of ($/KWh) for plots on the right side, while the y-axis represents the average Equity Index, which measures the equitable distribution of charging infrastructure. The blue line represents disadvantaged census tracts, and the orange line represents non-disadvantaged census tracts.

For Level 1 charging stations, we observe that changes in charging speed and cost have a negligible effect on the average accessibility index for both community types. This reflects the nature of Level 1 charging, which is slow and typically utilized for overnight charging, showing almost no impact on immediate charging station accessibility.

In contrast, the sensitivity of the average accessibility index to the charging speed of DC fast chargers (alpha 3) shows an initial increase with higher miles of range per hour, followed by a decrease beyond a certain threshold. This illustrates that while the availability of fast charging can enhance accessibility, there is a point beyond which further speed increases do not translate into proportional accessibility gains. The trend is evident in both community types, with a more pronounced decline in disadvantaged tracts, illustrating diminishing returns on further investments in charging speed alone. This trend is almost the same for level 2 charging stations as well for both disadvantaged and non-disadvantaged communities.

Overall, the results of sensitivity analysis indicate that while the deployment of faster chargers is crucial, improving accessibility equitably necessitates addressing additional factors, including the strategic distribution of charging stations and consideration of the total number of stations, to ensure that enhancements in charging infrastructure benefit all segments of the population.

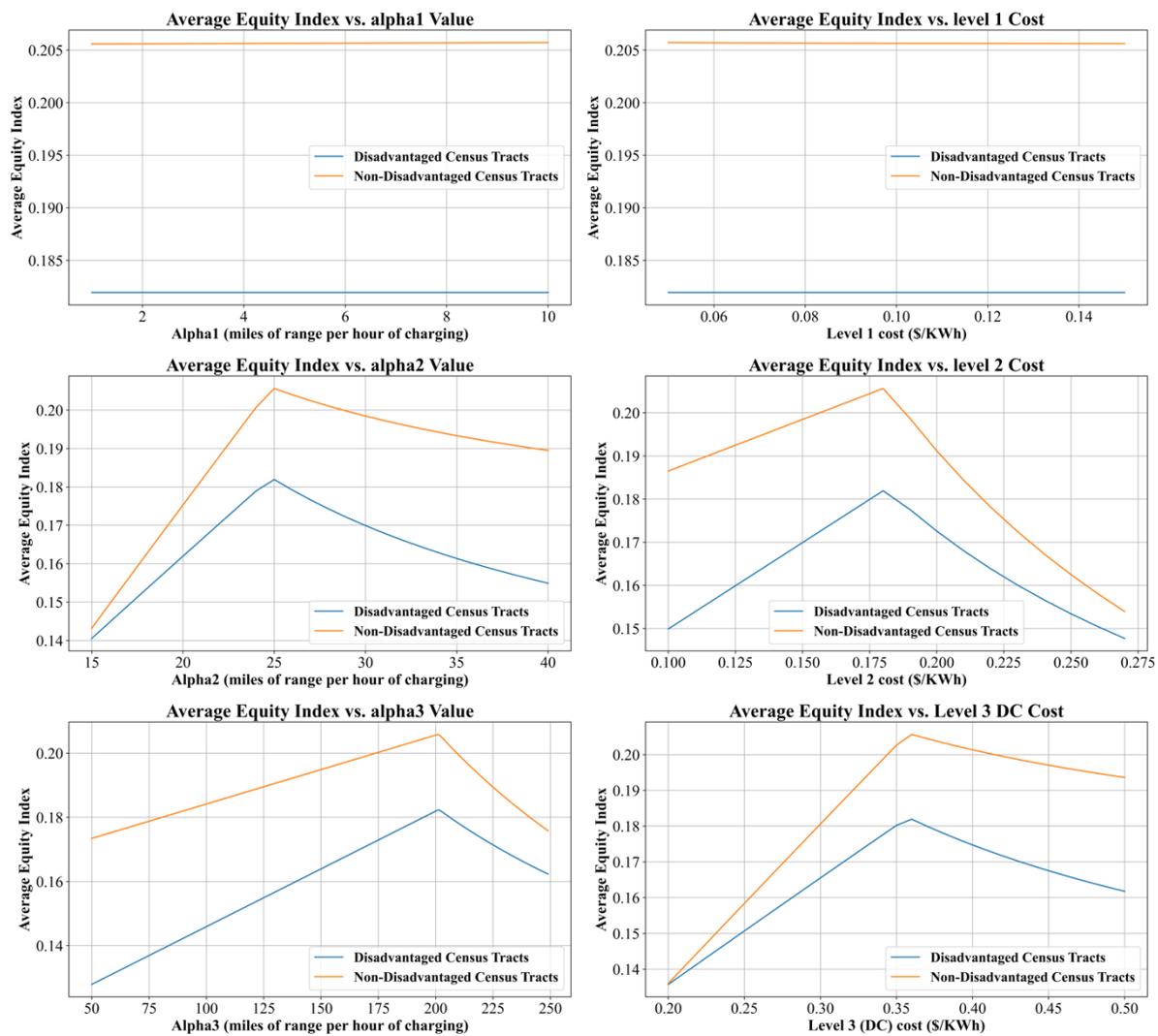

**Figure 4. Sensitivity analysis of the average equity index in response to varied charging speeds and costs for Level 1, Level 2, and DC Fast charging stations across Disadvantaged and Non-Disadvantaged census tracts**

**Strategies for increasing equity**

The sensitivity analysis results in *Figure 5* indicate that the approach to increasing electric vehicle charging infrastructure can significantly affect the average accessibility within census tracts. While adding an identical number of charging stations to each tract has a negligible impact on accessibility, accessibility significantly increases when the number of stations is proportionate to the population of each tract. This demonstrates the importance of considering population density and the unique characteristics of each census tract when planning infrastructure improvements. Based on the findings, it is recommended to perform population-proportionate distribution of charging stations, as this approach significantly enhances accessibility. Consequently, this strategy maximizes the charging station network to facilitate greater adoption of electric vehicles.

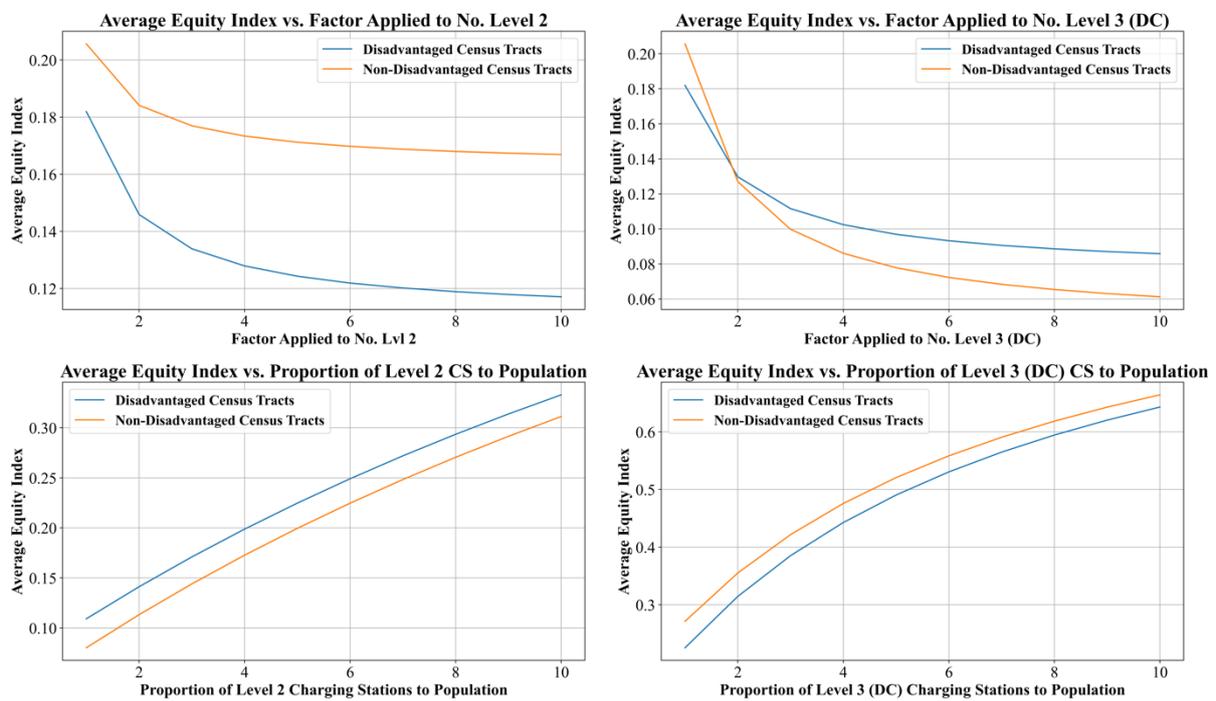

**Figure 5. Impact of scaling charging station numbers on average equity index in Disadvantaged and Non-Disadvantaged census tracts**

## 6. CONCLUSION

This research endeavor has culminated in the development of a generalized framework for conducting equity analysis within ground transport infrastructure. Addressing the historical disparities in transportation infrastructure that have perpetuated unequal access and opportunities for different groups is of paramount importance. Acknowledging and redressing these inequities can promote justice and equal access to essential resources such as jobs, education, and healthcare. The proposed framework outlines a multi-step process for assessing equity in transportation infrastructure projects. It begins with identifying who will be impacted by the project, collecting relevant socio-economic data, and identifying disadvantaged groups. Crucial to this process is establishing equity criteria that align with project goals. These criteria are essential for ensuring that the infrastructure serves all segments of the population equitably, particularly those who have been historically underserved or marginalized. The next step is to determine the equity analysis approach based on the specific context of the project and the availability of data and resources and conduct the equity analysis. The final step involves documenting and monitoring the equity criteria over time to understand the evolving impacts of the transportation infrastructure on different communities.

In a practical application, the framework was used to assess the equity impact of EV chargers in Baltimore City, Maryland, through a meso-scale case study accompanied by a sensitivity analysis. Surprisingly, some socioeconomically disadvantaged communities showed high availability of charging infrastructure, while some non-disadvantaged tracts were found to have severe shortages. The results highlight the requirement for a standardized approach to equity analysis, ensuring precise and equitable transportation development for future endeavors. This discrepancy underscores the existence of infrastructure gaps across various socio-economic contexts. While this study makes a significant contribution to understanding and addressing equity considerations in transportation infrastructure, there is an ongoing need for standardizing equity analysis approaches and developing comprehensive measures for accurately assessing the equity impact of transportation projects.

**DATA AVAILABILITY STATEMENT:**

Some or all data, models, or codes that support the findings of this study are available from the corresponding author upon reasonable request.